\documentstyle[graphicx,times]{mn}

\title{A  CCD Study of High Latitude Galactic Structure: Testing the Model
Parameters}
\author[Karaali et al.]
       {S.~Karaali,$^{1,2}$ S.~G.~Ak,$^1$ S.~Bilir,$^1$ Y.~Karata\c{s}$^1$, and G.~Gilmore$^3$\\
  $^1$Istanbul University Science Faculty, Department of Astronomy and Space
      Sciences, 34452 University-Istanbul, Turkey\\
  $^2$Visiting Astronomer, Institute of Astronomy, Madingley Road, Cambridge, CB3 OHA, UK\\
  $^3$Institute of Astronomy, Madingley Road, Cambridge, CB3 OHA, UK\\}

\date{Accepted 2002 month day.
      Received year month day;
      }

\pagerange{\pageref{firstpage}--\pageref{lastpage}}

\begin{document}

\maketitle

\label{firstpage}

\begin{abstract}
We interpret published CCD {\it UBVI\/} data to deduce the stellar
density distribution and metallicity distribution function in the
region from $2-8$ kpc from the Galactic Plane, and compare our results to
several star count models. A feature of extant star count models is
degeneracy between the adopted scale heights of the thin and thick
disks, and their local normalisation. We illustrate the utility of
this small data set, and future larger sets (e.g. {\it SDSS\/}), by explicitly
considering consistency between the derived density laws, and the
implied solar neighbourhood luminosity function. Our data set, from
Hall et al.\ (1996) ($l=52^{o}, b=-39^{o}$) contains 566 stars, selected to be
consistent with stellar loci in colour-colour diagrams. The effective
apparent {\it V\/}-magnitude interval is $15.5\leq V_{o}\leq 20.5$. Our analysis
supports the parameterisation of the recent ({\it SDSS\/}) galaxy model of
Chen et al.\ (2001), except in preferring the stellar halo axis ratio to
be $\eta=0.84$.
 
Photometric metal-abundances have been derived for 329 stars with
${\it (B-V)_{o}\/}\leq1.0$ using a new calibration. This show a multimodal
distribution with peaks at $[Fe/H]$=-0.10, -0.70, and -1.50 and a tail
down to -2.75 dex. The vertical distance-dependent metallicity distribution function, if
parameterised by a single mean value, can be described by
a metallicity gradient $d[Fe/H]/dz\sim-0.2$ dex/kpc for the thin disk and
thick disk, and $d[Fe/H]/dz\sim-0.1$ dex/kpc for the inner halo, to
$z$=8 kpc. The data are however better described as the sum of three
discrete distribution functions, each of which has a small or zero
internal gradient. The changing mix of thin disk, thick disk and halo
populations with distance from the plane generates an illusion of a
smooth gradient.

\end{abstract}


\section{Introduction}

The traditional star count analyses of Galactic structure have provided a
picture of the basic structural and stellar populations of the Galaxy.
Examples and reviews of these analyses can be found in Bahcall\ (1986),
Gilmore et al.\ (1989), Majewski\ (1993), Robin et al.\ (2000), and 
recently in Chen et al.\ (2001). The largest of the observational
studies prior to {\it SDSS\/} are based on  photographic surveys; 
the Basle Halo Program (Becker\ 1965) has presented
the largest systematic photometric survey of the Galaxy (Fenkart\ 1989a,
b, c, d; Del Rio \& Fenkart\ 1987; Fenkart \& Karaali\ 1987;  Fenkart \&
Karaali\ 1990; Fenkart \& Karaali\ 1991). The Basle Halo Program photometry
is currently being recalibrated and reanalysed, using an improved
calibration of the RGU photometric system (Buser et al.\ 1998, 1999; Ak et
al.\ 1998; Karata\c{s} et al.\ 2001). More recent and future studies are being
based on CCD survey data. Most have in general much smaller area coverage or a
restriction to only high Galactic latitudes, or a focus at faint
magnitudes (e.g. Willman et al.\ 2002; Chen et al.\ 2001). HST studies are 
a limiting case (e.g. Johnson et al.\ 1999), with very deep but extremely 
small area coverage, and corresponding very poor statistical weight. The 
general absence of CCD {\it UV\/} data additionally makes such analyses 
sensitive to assumptions on metallicity distributions.

Even small-area CCD studies probing intermediate apparent magnitudes can be
valuable, however, when analysed in the light of known solar neighbourhood
constraints, especially consistency with the local stellar luminosity
function. This is required since star-count analyses are essentially an 
attempt to deconvolve the product of a density profile and a local
normalisation, with that local normalisation being the solar neighbourhood 
luminosity function for the specific stellar population of relevance. The 
local luminosity function determined from Hipparcos parallax data
and that deduced from star count analyses must be consistent, providing an
additional constraint on Galactic modelling, or an independent check on 
photometric calibrations. Here we illustrate this consistency by analysing
the small area 5-colour {\it UBVRI\/} CCD survey by Hall et al.\ (1996).

We do not here use the available SDSS data, partly since these have
recently been analysed by Chen etal, but alos since our aim is to
illustrate the general approach. Over the next few months massive
photometric data sets will become available from SDSS, 2MASS, DENIS,
UKIDSS, VST, CFH/Megacam, Suprime,.... These data sets will combine
statistical weight with the wide area coverage which will allow
consideration of second-order effects, quantifying the structure of
the Galaxy beyond simple analytic smooth functions. Given that, it is
timely now to consider method, rather than specific interim results.

In addition to a direct test of the density profile/luminosity
function consistency requirements, we use the information available in the
multi-colour photometry, especially the {\it U\/}-band data, to derive 
limits on metallicity gradients in the thin disk, thick disk, and halo.

The existence of a clear vertical metallicity gradient for any
pressure-supported component of
the Galaxy means that it formed by dissipative collapse. The pioneers of
this suggestion are Eggen, Lynden-Bell, \& Sandage\ (1962 hereafter ELS) 
who argued that the Galaxy collapsed in a free-fall time ($\sim 2x10^{8}$ yr).
A discussion of the current status of this model is provided by Gilmore,
Wyse \& Kuijken\ (1989). Over the past 20 years, observational studies have 
revealed that the collapse of the Galaxy occured slowly with the limiting 
case being assembly of the Galaxy on many dynamical times, which (now allowing 
for a dark matter halo), implies times of very many Gyr (e.g. Yoshii \& Saio\
1979; Norris, Bessel \& Pickles\ 1985, hereafter NBP; Norris\ 1986; Sandage \&
Fouts\ 1987; Carney, Latham \& Laird\ 1990; Norris \& Ryan\ 1991; Beers \&
Sommer-Larsen\ 1995). This picture was postulated largely on a supposed wide 
age in the globular cluster system (Searle \& Zinn\ 1978, hereafter SZ; 
Schuster \& Nissen\ 1989). SZ especially argued that the Galactic halo was 
not formed in an ordered collapse, but from merger or accretion of numerous 
fragments, such as dwarf-type galaxies. Such a scenario indicates no 
metallicity gradient or young and even more metal-rich objects at the 
outermost part of the Galaxy. The globular cluster age range supposition has 
been disproved by recent analyses (Rosenberg et al.\ 1999), while the number 
of young field halo stars has been shown to be extremely small, inconsistent 
with the model, by Unavane, Wyse \& Gilmore\ (1996) and Preston \& Sneden\ 
(2000, see also Gilmore\ 2000). Nonetheless, hierachical models have become 
the default (Freeman \& Bland-Hawthorn\ 2002). We readdress the metallicity 
gradient from the present data.

In Section 2, we describe the selection of the sample (566 stars) from
the 4462 objects (stars, galaxies, quasars etc.) observed by Hall et
al.\ (1996), separation of the sample stars into different populations,
and their absolute magnitude determination. Section 3 discusses the
density functions, for seven absolute magnitude intervals evaluated
for distances beyond $r=0.4$ kpc, with three galactic models and
comparison of the resulting luminosity functions with that of
Hipparcos (Jahreiss \& Wielen\ 1997) and Gliese \& Jahreiss\ (1992). In
Section 4, we search for a metallicity gradient in each of the
galactic components, i.e.: thin disk, thick disk, and halo. Section 5
provides a summary and discussion.  Finally, a summary is given for
the new calibration in an appendix.

\begin{figure}
\caption{Two-colour diagrams for 922 objects with stellar image
profiles from Hall et al.\ (1996). (a) for $(U-B)_{o}$ versus $(B-V)_{o}$, and 
(b) for $(V-I)_{o}$ versus $(B-V)_{o}$. Objects with $(U-B)_{o}<-0.46$ mag,
corresponding to $(u'-g')_{o}<-0.5$ mag in the Sloan photometry are
extra-galactic objects}
\end{figure}

\begin{figure}
\caption{Two-colour diagrams for objects with $(U-B)_{o}\geq-0.46$ and
$V_{o}\leq20.5$. (a) for $(U-B)_{o}$ versus $(B-V)_{o}$, and (b) for $(V-I)_{o}$ versus
$(B-V)_{o}$. The $U-B$ selection has reduced the scatter relative to that in
Fig.1a and b.}
\end{figure}

\section{Data}
\subsection {Star-Galaxy-QSO separation}

The data are taken from the catalogue of Hall et al. (1996), who provide a
deep multi-colour survey for 12 CCD fields. From the 4462 sources they
detected, we selected the  stellar-like sources labelled with "s", "sf", "Fs"
in the catalogue for Field 21e-w (Galactic coordinates $l=52^{o}, b=-39^{o}$,
area 0.149 square degree). We adopted the mean of two $E(B-V)$ colour-excesses
given by Hall et al.\ (1996), i.e.: 0.0223 mag, for all stars and we de-reddened
{\it U-B\/} and {\it V-I\/} colour-indices by the following well known equations:

\begin{equation}
E(U-B) = 0.72E(B-V) + 0.05E^{2}(B-V)
\end{equation}

\begin{equation}
E(V-I) = 1.250[1+0.06 (B-V)_{o} + 0.014E(B-V)]E(B-V) 
\end{equation}
the total absorption $A(V)$ is evaluated as usual,\\ i.e.: $A(V)=3.1 E(B-V)$.

We restricted the sample at the faint end, corresponding to the peak
in the distribution of apparent magnitudes, at $V_{o}=20.5$ mag, leaving
922 sources. The two-colour diagrams $(U-B)_{o}$ - $(B-V)_{o}$ and
$(B-V)_{o}$ - $(V-I)_{o}$, for these objects indicate residual
significant contamination by extra-galactic objects (galaxies and QSO) and,
WD and BHB stars (Fig.1a and b). We then rejected all sources with 
$(U-B)_{o}<-0.46$, which corresponds to the location of the bluest 
extra-galactic objects, $(u'-g')_{o}<-0.5$ mag, in {\it SDSS\/} 
(Chen et al.\ 2001). This $(U-B)$ cut removed most outliers in the 
$(BVI)$ two-colour diagrams (Fig.2). We further removed those few
sources which lay significantly off the stellar locus in $(UBVI)$, 
limiting the sample to 566 sources with stellar colours (Fig.3).

\begin{figure}
\caption{Two-colour diagrams for the final sample after excluding outliers in Fig.2a 
and b. (a) for $(U-B)_{o}$ versus $(B-V)_{o}$, and (b) for $(V-I)_{o}$ versus
$(B-V)_{o}$.}
\end{figure}

\subsection {Stellar population types and absolute magnitude determination}

The $(B-V)_{o}$ colour distribution of the sample stars shows a bimodal
distribution (Fig.4), as expected for a high-latitude field (see, e.g.
Phleps et al.\ 2000). In order to assign (statistical) distances, we need
to distinguish (statistically) between the three basic metallicity-dependent
populations, thin disk, thick disk and halo. It is well known 
(cf Chen et al.\ (2001) for a recent discussion) that population types are a 
complex function of both colour and apparent magnitude. According to
Chen et al., the halo has a turnoff at $(g'-r')_{o}=0.20$ mag and it dominates
in the apparent magnitude interval fainter than $g'_{o}\sim 18$ mag, whereas the
thick disk has a turnoff at $(g'-r')_{o}=0.33$ mag and it is dominant at 
brighter apparent magnitudes, $g'_{o} < 18$ mag. The corresponding turnoff
colours in the $UBVRI$ system are $(B-V)_{o}=0.41$ and $0.53$ mag for halo and
thick disk, respectively. The diagram $V_{o}$ versus $(B-V)_{o}$ in Fig.5 reveals
these three populations, with, for example, the blue shift of the turnoff
moving from thick disk to halo being apparent near $V=18$. It seems that thick
disk and halo populations overlap in Fig.4. Hence, the colour distribution of
the sample stars is given as a function of apparent magnitude (Fig.6) and the
turnoffs for thick disk and halo, as well as for thin disk, are fixed 
precisely (Table 1).

\begin{table}
\centering
\caption{The colour-magnitude intervals most appropriate for
statistical discrimination of the three stellar populations.}
\begin{tabular}{cccc}
\hline
Populations & Thin Disk & Thick Disk & Halo\\
$V_{o}$& $(B-V)_{o}$ & $(B-V)_{o}$ & $(B-V)_{o}$\\
\hline
(15.5-16.0] & $\geq 0.9$ & $<0.9$ & ---\\
(16.0-17.0] & $\geq 0.9$ & $<0.9$ & ---\\
(17.0-18.0] & $\geq 0.9$ & $<0.9$ & ---\\
(18.0-19.0] & $\geq 1.0$ & $[0.5-1.0)$ & $<0.5$\\
(19.0-20.0] & $\geq 0.9$ & --- & $<0.9$\\
(20.0-20.5] & $\geq 0.9$ & --- & $<0.9$\\
\hline
\end{tabular}
\end{table}

\begin{figure}
\caption{Colour distribution for 566 stars in our sample. The thin disk
population, $(B-V)_{o}>1.0$ mag, is rather conspicuous whereas thick disk and
halo populations, $(B-V)_{o}<1.0$, overlap.}
\end{figure}

\begin{figure}
\caption{$V_{o}$ versus $(B-V)_{o}$ diagram for the sample stars. Contrary to the
distribution in Fig.4, thick disk and halo stars can be distinguished,
with, for example, the appearance of a blue halo turnoff being apparent near
$V_{o}=17.5$, $(B-V)_{o}=0.4$.}
\end{figure}

\begin{figure*}



\caption{Colour distributions for 566 stars in our sample as a function of
apparent magnitude. The colour limits revealed from these panels
which allow sub-population isolation are given in Table 1.}
\end{figure*}

\begin{figure*}

\caption{$M(V)-(B-V)_{o}$ colour-absolute magnitude relations for three
populations. (a) for thin disk, (b) for thick disk, and (c) and (d) for 
halo (see text for details).}
\end{figure*}

Assignment of individual absolute magnitudes for stars classified into
the different populations are determined by means of appropriate
colour-absolute magnitude relations, as follows. The $M(V)$ absolute
magnitudes and $(B-V)_{o}$ of Lang\ (1992) are used to define the
colour-absolute magnitude relation for thin disk stars (Fig.7a). The
colour-absolute magnitude relation for thick disk stars is adopted
from the globular cluster 47 Tuc ($[Fe/H]=-0.65$ dex), with data taken
from Hesser et al.\ (1987). These authors provide $V$ and $B-V$ data
as well as $E(B-V)$ colour excess ($0.04$ mag) and apparent distance
modulus, $V-M(V)=13.40$ mag, which gives the absolute magnitude of a
star by combination with the total absorption, $A(V)=3.1 E(B-V)$
(Fig.7b). For the halo, we used two colour-absolute magnitude
relations, that for the globular cluster M13 ($[Fe/H]=-1.40$ dex) for
stars with $(B-V)_{o}\geq 0.40$ mag and that of the globular cluster
M92 ($[Fe/H]=-2.20$ dex) for only an interval less than one magnitude
which is not covered by the diagram of M13 ($0.30<(B-V)_{o}\leq
0.40$).  We applied the same procedure as noted above for the
calibration of 47 Tuc, to the data of Richer \& Fahlman\ (1986) and
Stetson \& Harris\ (1988). Richer \& Fahlman provide $E(B-V)=0.02$ mag
and $V-M(V)=14.50$ mag for M13, while Stetson \& Harris give
$E(B-V)=0.02$ mag and $V-M(V)=14.60$ mag for M92. The colour-absolute
magnitude relations for M13 and M92 are given in Fig.7c and d.

\begin{figure*}

\caption{Comparison of logarithmic space densities with the galactic
model of BRK for (a) $4<M(V)\leq5$, (b) $5<M(V)\leq6$, (c) $6<M(V)\leq7$, 
(d) $7<M(V)\leq8$, (e) $8<M(V)\leq9$, (f) $9<M(V)\leq10$, and (g) $10<M(V)\leq11$ 
mag. Heavy dots designate the centroid-distance
$r^{*}=[(r^{3}_{1}+r^{3}_{2})/2]^{1/3}$ of the corresponding partial volume 
$\Delta V_{1,2}$.}
\end{figure*}

\begin{figure}
\caption{The stellar luminosity functions, at $r=0$ kpc, resulting from comparisons 
of derived space densities with galactic models, (a) Gilmore \& Wyse (1985), (b) 
Buser, Rong \& Karaali\ (1999), and (c) Chen et al. (2001), compared to that of 
Hipparcos (Hip 1997), and Gliese \& Jahreiss\ (GJ 1992).}
\end{figure}

\section{Density Functions and Luminosity Function}

The logarithmic space densities $D^{*}=log D+10$ are evaluated for stars for
seven absolute magnitude intervals, i.e.: $4<M(V)\leq5$, $5<M(V)\leq6$, 
$6<M(V)\leq7$, $7<M(V)\leq8$, $8<M(V)\leq9$, $9<M(V)\leq10$, and $10<M(V)\leq11$ 
mag, over the distance range for which absolute magnitudes are completely sampled 
by the available photometry (Table 2). The number of stars brighter than $M(V)=4$ 
is also given in the same table. Here, $D = N/\Delta V_{1,2}$, N being the number 
of stars, found in the  volume $\Delta V_{1,2}$, which is determined by its 
limiting distances $r_{1}$ and $r_{2}$ and by the apparent field-size in square 
degrees $\sq$, i.e.: ${\Delta}V_{\rm 1,2}=({\frac{\pi}{180}}) ^{2} ({\frac{\sq}{3}}) 
(r_{\rm 2}^{\rm 3}-r_{\rm 1}^{\rm 3})$.

\begin{figure}
\caption{The stellar luminosity function, at $r=0$ kpc, resulting from comparisons 
of derived space densities with the galactic model of Chen et al.\ (2001), with 
some modifications. In (a) and (b) the axis ratio for the halo is adopted as 0.65 
and 0.84 respectively, and in (c) the density law for the halo is assumed to be 
de Vaucouleurs instead of power-law (with axis ratio 0.84). Comparison of these 
luminosity functions with that of Hipparcos (Hip 1997), and Gliese \& Jahreiss 
(GJ 1992) favours the models in the lower two panels.}
\end{figure}

\begin{table*}
\caption[]{Logarithmic space densities, $D^{*} = log D+10$ for seven absolute magnitude intervals, where $D = N/ \Delta V_{1,2}$, N being the number of stars, found in the partial volume $\Delta V_{1,2}$, which is determined by its limiting distances $r_{1}$ and $r_{2}$ and by the apparent field - size in square degrees ${\sq}$; i.e.: ${\Delta}V_{\rm 1,2}=({\frac{\pi}{180}}) ^{2} ({\frac{\sq}{3}}) (r_{\rm 2}^{\rm 3}-r_{\rm 1}^{\rm 3})$. $r^{*}=[(r^{3}_{1}+r^{3}_{2})/2]^{1/3}$: centroid-distance of the partial volume $\Delta V_{1,2}$. Two thick horizontal lines for each absolute magnitude interval define the distance interval for completeness (distances in kpc, volumes in $pc^{3}$).}
\begin{center}
\begin{tabular}{cccccccccccc}
\hline
 &  & M(V)$\rightarrow$ & (2-3] & (3-4] & (4-5] & (5-6] & (6-7] & (7-8] & (8-9] & (9-10] & (10-11] \\
\hline
$r_{1}$-$r_{2}$& $\Delta V_{1,2}$ & r* &~N~~D*&~N~~D*&~N~~D*&~N~~D*&~N~~D*&~N~~D*&~N~~D*&~N~~D*&~N~~D*\\
\hline
~0.00-~0.40& 9.54(02)&~0.32&      &        &        &        &        &        &\underline {~1~~7.02}&\underline {~3~~7.50}&\underline {~1~~7.02}\\      
~0.40-~0.63& 2.85(03)&~0.54&      &        &        &        &        &$\overline {~2~~6.85}$&~6~~7.32&~9~~7.50&\underline {~7~~7.39}\\      
~0.63-~1.00& 1.13(04)&~0.86&      &        &        &~5~~6.64&$\overline {13~~7.06}$&~3~~6.42&17~~7.18&\underline {16~~7.15}&~5~~6.64\\
~1.00-~1.59& 4.51(04)&~1.36&      &        &~4~~5.95&22~~6.69&21~~6.67&14~~6.49&\underline {19~~6.62}&28~~6.79&~1~~5.35\\
~1.59-~2.51& 1.80(05)&~2.15&      &        &13~~5.86&$\overline {27~~6.18}$&44~~6.39&\underline {26~~6.16}&25~~6.14&~4~~5.35&        \\
~2.51-~3.98& 7.15(05)&~3.41&      &        &$\overline {13~~5.26}$&25~~5.54&\underline {27~~5.58}&19~~5.42&~3~~4.62&        &        \\
~3.98-~6.31& 2.85(06)&~5.40&      &        &17~~4.78&22~~4.89&18~~4.80&~4~~4.15&        &        &        \\
~6.31-~7.94& 3.78(06)&~7.22&      &~2~~3.72&~9~~4.38&\underline {~7~~4.27}&~4~~4.02&        &        &        &        \\
~7.94-10.00& 3.78(06)&~9.09&      &~3~~3.60&12~~4.20&~3~~3.60&        &        &        &        &        \\
10.00-12.59& 1.51(07)&11.44&~1~~--&~2~~3.12&\underline {10~~3.82}&~7~~3.67&        &        &        &        &        \\
12.59-15.85& 3.00(07)&14.40&      &~3~~3.00&~6~~3.30&        &        &        &        &        &        \\
15.85-17.78& 2.48(07)&16.87&      &        &~3~~3.08&        &        &        &        &        &        \\
\hline
           &         &Total&~1~~~~&10~~~~~~~&87~~~~~~~&118~~~~~&127~~~~~&68~~~~~~&71~~~~~~&60~~~~~~&14~~~~~~\\
\hline
\end{tabular}  
\end{center}
\end{table*}

\begin{table*}
\centering
\caption{Model parameters of Buser et al.\ (BRK 1998, 1999), Gilmore \& Wyse\
(GW 1985), and Chen et al.\ (C 2001) and their comparison (fifth and sixth
columns). Symbols: $n_{i}$ (i = 0, 1, 2, 3) local density relative to thin
disk, $H_{i}$ (i = 0, 1, 2) scale-height in pc, $h_{i}$ (i = 0, 1, 2) scale-lenght
in pc, $R_{eff}$: effective radius in pc, $R_{\sun}$ distance of the Sun to the
Galactic center in pc, $\eta$: axis ratio for halo, and $r_{c}$: core radius in
pc.}
\begin{tabular}{cccccc}
\hline
Authors & BRK & GW & C & BRK-GW & BRK-C\\
\hline
Thin Disk& Double Exponential&Double Exponential  &Double Exponential &--- & ---\\
$n_{0}$  & $0.2^{(1)}$       & 0.2                &$0.2^{(1)}$        &--- & ---\\
$n_{1}$  & 1.0~~~            & 1.0                &1.0~~~             &--- & ---\\
$H_{0}$  & 170~~~            & 100                &90 ~~~             &--- & 80 \\
$H_{1}$  & 292.5~            & 300                &330 ~~             &--- & ---\\
$h_{1}$  & 4010~~            & ~4000              &2250~              &--- & 1760\\
\hline
Thick Disk& Double Exponential&Double Exponential  &Double Exponential&--- & ---\\
$n_{2}$  & 0.059~~           & 0.02               &0.075~             &--- & ---\\
$H_{2}$  & 910~~~~           & ~1000              &750 ~~             &90  & 160\\
$h_{2}$  & 3000~~            & ~4000              &3500~              &--- & 500\\
\hline
Halo      & de Vaucouleurs    &de Vaucouleurs      &Power-law          &--- & ---\\
$n_{3}$   & 0.0005            & ~~0.001            &~~~0.00125         &--- & ---\\
$R_{eff}$ & 2696~~~           & ~2700              &--- ~~             &--- & ---\\
$\eta$    & 0.84~~~           & ~0.85              &0.55~              &--- &0.29\\
$R_{\sun}$& 8600~~            & ~8500              &8600~              &--- & ---\\
Power-law index& ---~~        & ~---               &2.5~~~             &--- & ---\\  
$r_{c}$   & ---~~             & ~---               &1000~              &--- & ---\\
\hline
$^{(1)}$ adopted
\end{tabular}
\end{table*}

The density functions are most conveniently presented in the form of
histograms whose sections with ordinates $D^{*}(r_{1}, r_{2})$ cover the
distance-intervals ($r_{1}$, $r_{2}$). Heavy dots on the histogram sections
$D^{*}(r_{1}, r_{2})$ designate the centroid-distance 
$r^{*}=[(r^{3}_{1}+r^{3}_{2})/2]^{1/3}$ of the corresponding partial volume
$\Delta$$V_{1,2}$ (Del Rio \& Fenkart 1987; Fenkart \& Karaali\ 1987; Fenkart\ 
1989a, b, c, d). The density functions are compared with three galactic models, 
i.e.: Gilmore \& Wyse\ (1985, hereafter GW); Buser, Rong, \& Karaali\ (1998, 
1999; hereafter BRK); and Chen et al.\ (2001, hereafter C) given in the form 
$\Delta logD(r)=log D(r,l,b)-logD(0,l,b)$ versus $r$, where $\Delta logD(r)$ is
the logarithmic difference of the densities at distances $r$ and at the Sun. Thus,
$\Delta logD(r)$=0 is the logarithmic space density at $r=0$, which is the 
parameter required for luminosity function determination. The comparison is carried
out as explained in several studies of the Basle fields (Del Rio \& Fenkart\ 1987; 
Fenkart \& Karaali\ 1987), i.e.: by shifting the model curve perpendicular to the
distance axis until the best fit to the histogram results at the centroid
distances. Fig.8 shows the comparison of the observed density functions
with the model BRK as an example.

There is adequate agreement between both models and the observed density functions
within the limiting distance of completeness marked by horizontal thick lines in
Table 2. However, this is not the case when one includes the luminosity functions.
As cited above, the luminosity function close to the Sun: $\varphi^{*}(M)$, i.e.: 
the logarithmic space density for the stars with M$\pm0.5$ mag at $r=0$ is 
the $D^{*}$-value corresponding to the intersection of the model-curve with the 
ordinate axis of the histogram concerned. The luminosity functions resulting from
comparisons of our space density data with the models GW, BRK, and C confronted to 
the luminosity function of Hipparcos (Jahreiss \& Wielen\ 1997) and that of 
Gliese \& Jahreiss\ (GJ 1992) are given in Fig.9a, b, and c, respectively.

When we compare the luminosity functions obtained in this work with the luminosity
function from Hipparcos, the Buser et al. (BRK) model is most successful for 
intrinsically bright stars, $M(V)\leq8$ mag, whereas the model of Chen et al. fits
the data better for intrinsically fainter stars, $M(V)>8$ mag. Obviously, the 
reason for the difference in this comparison of effective local luminosity functions 
is due to the difference between the parameters used (Table 3). For the models of
Gilmore \& Wyse (GW) and of Buser et al. (BRK), the main difference is between the
scale-height and local density of the thick disk, whereas for Buser et al. (BRK) 
and Chen et al. (C) the differences involve five parameters, i.e.: the scale-heights
and scale-lengths of the thin and thick disks, and the axis ratio of the halo. The  
effect of the different density law (power-law) used for the halo in the model of
Chen et al. will be discussed below.

We modified the C-model of Chen et al. by changing the halo axis ratio from their 
adopted 0.55 to 0.65 and 0.84, respectively. The luminosity function resulting from 
comparison of this modified model with the observed density functions is in substantially 
improved agreement with the luminosity function of Hipparcos. This modified model matches 
the data better overall than does the Buser et al. BRK model (Fig.10a and b). 
Now, a question arises from this comparison whether or not a power-law for the 
halo density matches the observations to the Galactic models better. We therefore 
recalculated the Chen et al. model, adopting a de Vaucouleurs spheroid denity law
(with axis ratio 0.84) for the halo in place of the Chen et al. power-law. Comparison of 
this new model with the local normalisation data (Fig.10c) shows an improved fit, relative 
to the power-law model. Hence, regarding the best fit of the local luminosity function
constraint resulting from comparison of the observed density functions for absolute
magnitude intervals $4<M(V)\leq5$, $5<M(V)\leq6$, $6<M(V)\leq7$, $7<M(V)\leq8$, 
$8<M(V)\leq9$, $9<M(V)\leq10$, and $10<M(V)\leq11$, we conclude that the data 
suggest an increase in the axis ratio in the density law for the halo, to a value 
of $\eta=0.84$, and further slightly prefer a halo density profile described by a
de Vaucouleurs profile rather than a power-law.

\begin{figure*}
\caption{Metallicity distribution for stars with $(B-V)_{o}\leq1.0$ mag as a function 
of apparent magnitude $V_{o}$ (panels a-f), and for their combination within the 
limiting apparent magnitude, $V_{o}\leq20.5$ (last panel). (a) $15.5<V_{o}\leq16$, 
(b) $16<V_{o}\leq17$, (c) $17<V_{o}\leq18$, (d) $18<V_{o}\leq19$, (e) $19<V_{o}\leq20$, 
(f) $20<V_{o}\leq20.5$, and (g) $15.5<V_{o}\leq20.5$. Curves in (g) are the fitted 
gaussians distributions for three populations (continuous curve) and their sum 
(dashed curve).}
\end{figure*}

\section{Metallicity Distribution}

The metal abundances for 329 stars with $(B-V)_{o}\leq 1.0$ mag were evaluated by
means of a new calibration, of the standard metallicity-dependent ultraviolet-excess 
photometric parameter $\delta_{0.6}$, i.e.: 
$[Fe/H]=0.10-2.76\delta-24.04\delta^{2}+30.00\delta^{3}$, obtained via 88 dwarfs
where the determination of abundances for most of them is based on high-resolution 
spectroscopy (Karaali et al.\ 2002). The metallicity distribution for the sample of 
all stars is multimodal (Table 4 and Fig.11g); one sees three local maxima, at 
$[Fe/H]=-0.10$, $-0.70$, and $-1.50$ dex, and a tail down to $-2.75$ dex. 

However, one notices a systematic shift from the metal-rich stars to the metal-poor
ones, when the distribution is considered as a function of apparent magnitude 
(Fig.11a-f). This is particularly apparent in Fig.12, where the mean metallicity 
as a function of $z$-distance is displayed. The overall distribution shows a 
continuous metallicity gradient $d[Fe/H]/dz=-0.20$ dex/kpc, up to $z=8$ kpc. It is
interesting that the gradient is only marginally different for the thin disk 
($z<1.5$ kpc) and thick disk ($1.5<z<5$ kpc), whereas the halo shows a weak, if not 
zero metallicity gradient between 5 and 8 kpc, i.e.: $d[Fe/H]/dz=-0.10$ dex/kpc, 
and zero at larger distances. At face value this indicates a continuous smooth 
vertical abundance gradient through the thick disk. However, this presentation 
assumes that a single parameter, the mean, is adequate to describe a distribution 
function which is not gaussian, but is multi-modal. Is a single parameter a valid 
description of the data?

\begin{table*}
\centering
\caption{Metallicity distribution for 329 stars of all apparent magnitudes
(column 4) and for individual apparent magnitude intervals columns (5 -10), and the
corresponding modes.}
\begin{tabular}{ccccccccc}
\hline
&$V_{o}\rightarrow$&(15.5-20.5]&(15.5-16.0]&(16-17]&(17-18]&(18-19]&(19-20]&(20.0-20.5]\\
\hline
[Fe/H](dex)  & $<[Fe/H]>$(dex)& N & N & N & N & N & N & N \\
\hline
$(-3.0)-(-2.8)$&$-2.9$          &   &   &   &   &   &   &   \\
$(-2.8)-(-2.6)$&$-2.7$          &~1 &   &   &   &   &~1 &   \\
$(-2.6)-(-2.4)$&$-2.5$          &~2 &   &   &~1 &~1 &   &   \\
$(-2.4)-(-2.2)$&$-2.3$          &~5 &   &   &   &~3 &~2 &   \\
$(-2.2)-(-2.0)$&$-2.1$          &~9 &   &~1 &~1 &~1 &~4 &~2 \\
$(-2.0)-(-1.8)$&$-1.9$          &13 &   &   &   &~3 &~5 &~5 \\
$(-1.8)-(-1.6)$&$-1.7$          &14 &   &~1 &~1 &~5 &~6 &~1 \\
$(-1.6)-(-1.4)$&$-1.5$          &18 &   &   &~4 &~2 &~5 &~7 \\
$(-1.4)-(-1.2)$&$-1.3$          &11 &   &~1 &~1 &~5 &~3 &~1 \\
$(-1.2)-(-1.0)$&$-1.1$          &28 & ~1&~2 &~5 &~8 &~8 &~4 \\
$(-1.0)-(-0.8)$&$-0.9$          &28 & ~3&~2 &~3 &~7 &10 &~3 \\
$(-0.8)-(-0.6)$&$-0.7$          &39 & ~5&~5 &~9 &~9 &~7 &~4 \\
$(-0.6)-(-0.4)$&$-0.5$          &21 & ~2&~4 &~6 &~5 &~3 &~1 \\
$(-0.4)-(-0.2)$&$-0.3$          &43 & ~2&10 &12 &14 &~4 &~1 \\
$(-0.2)-(~~0.0)$&$-0.1$         &50 & ~6&12 &13 &10 &~6 &~3 \\
$(~~0.0)-(+0.2)$&+0.1           &47 & ~6&14 &11 &11 &~5 &   \\
\hline
             & total             &329& 25&52 & 67& 84& 69& 32\\
\hline
             & mode 1      &$-0.06$&$~~0.00$&$+0.03$ & $-0.13$&$-0.26$& $-0.07$&$--$\\
             & mode 2      &$-0.83$&$-0.72$&$-0.65$ & $-0.67$&$-0.73$& $-0.92$&$--$\\
             & mode 3      &$-1.59$& $--$&$--$  & $-1.50$&$-1.72$& $-1.70$&$--$\\
\hline
\end{tabular}
\end{table*}

To consider this in more detail, the modes are evaluated (Table 4) for the 
metallicity distribution in figures 11a-11g, and the metallicity distributions 
are given for different $z$-intervals, $z$ being the distance of a star to the 
Galactic plane in Table 5. The dips in Fig.11g separating three populations are 
statistically significant, for Hall et al.\ (1996) state that the external errors 
in their photometry as estimated from the two independent measurements of the 
magnitudes of each object, have been shown to be consistent with the internal errors 
computed according to photon statistics, except for an $\sim 2\%$ additional 
uncertainty independent of magnitude. This independent check proves that the 
flat-fielding process, aperture correction procedures, and photometry methods are all 
quite reliable, having inherent limitations of only the aformentioned $\sim 2\%$. 
As for systematic errors, their stellar locus matches values for stellar colours 
from the literature to about $5\%$. Three modes at $[Fe/H]=$ $-0.06$, $-0.83$, and $–-1.59$ 
dex for the distribution in Fig.11g correspond to the mean metal abundance for three 
components of the Galaxy, i.e.: thin disk, thick disk, and halo, though the one for 
the thick disk is a bit lower then the canonical one, $[Fe/H]=-0.65$ dex, probably 
affected by the metal poor tail of the thick disk (Norris\ 1996, see section 5 for 
detail). The gaussians fits with the modes just cited and their sum are also shown 
in Fig.11g.

As figures 11a-11f makes clear, the apparent abundance gradient is evidently an artefact 
of the changing relative proportions of the three populations present, thin disk, thick 
disk, and halo, with each population having no significant gradient. Each abundance 
distribution is simply consistent with a sum of three discrete distributions, with no 
systematic change in the mode of each. This suggestion can be confirmed by the modes for 
individual apparent magnitude intervals (Table 4) which show fluctuations with exception 
the mode for the thick disk for the apparent magnitude interval $19<V_{o}\leq20$, which is 
$\sim -0.2$ dex lower than the ones for brihter apparent magnitude intervals. This 
determination, with independent high-quality data and a new much improved photometric 
calibration, is essentially in agreement with the conclusions of Gilmore \& Wyse\ 
(GW 1985): the Galactic disks are better described as the sum of independent well-mixed 
sub-populations with different spatial distributions than as a continuum. However, the 
mean metal abundance in Table 5 show a systematic decrease with increasing mean $z$, 
indicating a slight vertical metallicity gradient for thin disk, thick disk, and inner 
halo (see section 5 for detail).

\begin{figure}
\caption{Mean metal abundance versus mean $z$-distance for 10 $z$-intervals, suggesting 
a metallicity gradient $d[Fe/H]/dz\sim-0.2$ dex/kpc for the thin disk and thick disk, 
and $d[Fe/H]/dz\sim-0.1$ dex/kpc for the inner halo. As shown in the text, this apparent 
smooth abundance gradient is an artefact of a mix of three independent distributions.}
\end{figure}

\begin{table*}
\centering
\caption{Metallicity distribution for 329 stars for 10 $z$-distance intervals, 
$z$ being the distance to the Galactic plane in kpc. Mean metal abundances and
 mean $z$ distances, as well as mean errors for the metallicity are also indicated.}
\begin{tabular}{cccccccccccc}
\hline
 &$z$(kpc)$\rightarrow$&(0-1]&(1-2]&(2-3]&(3-4]&(4-5]&(5-6]&(6-7]&(7-8]&(8-9]&(9-10]\\
\hline
[Fe/H](dex)  & $<[Fe/H]>$(dex)& N & N & N & N & N & N & N& N & N & N \\
\hline
$(-3.0)-(-2.8)$&$-2.9$          &   &   &   &   &   &   &   &   &   &    \\
$(-2.8)-(-2.6)$&$-2.7$          &   &   &   &   &~1 &   &   &   &   &    \\
$(-2.6)-(-2.4)$&$-2.5$          &   &   &   &   &   &~1 &   &   &   &    \\
$(-2.4)-(-2.2)$&$-2.3$          &   &   &~2 &   &~1 &   &   &~2 &   &    \\
$(-2.2)-(-2.0)$&$-2.1$          &   &   &~1 &~2 &~1 &~1 &~2 &   &~1 &~1  \\
$(-2.0)-(-1.8)$&$-1.9$          &   & ~1&~2 &~1 &~1 &~1 &~3 &~3 &   &~1  \\
$(-1.8)-(-1.6)$&$-1.7$          &   & ~2&~2 &~1 &~3 &   &~3 &~1 &~1 &    \\
$(-1.6)-(-1.4)$&$-1.5$          &~1 & ~3&~5 &~3 &~2 &   &~1 &~2 &   &    \\
$(-1.4)-(-1.2)$&$-1.3$          &   & ~3&~1 &~2 &~2 &   &~1 &~1 &~1 &    \\
$(-1.2)-(-1.0)$&$-1.1$          &~5 & ~5&~5 &~3 &~2 &~3 &~2 &~1 &   &    \\
$(-1.0)-(-0.8)$&$-0.9$          &~2 & ~8&~5 &~4 &~2 &~1 &~3 &~2 &~1 &    \\
$(-0.8)-(-0.6)$&$-0.7$          &~5 & 15&10 &~2 &~2 &~1 &   &~1 &   &~1  \\
$(-0.6)-(-0.4)$&$-0.5$          &~4 & ~5&~8 &~3 &   &~1 &   &   &   &    \\
$(-0.4)-(-0.2)$&$-0.3$          &10 & 17&~7 &~4 &~3 &   &   &   &   &    \\
$(-0.2)-(~~0.0)$&$-0.1$         &13 & 23&~8 &~4 &~2 &   &   &   &   &    \\
$(~~0.0)-(+0.2)$&+0.1           &15 & 25&~5 &~2 &   &   &   &   &   &    \\
\hline
             & total             &55  &107   &61   &31   &22& ~9& 15& 13& ~4& ~3 \\
\hline
             & $<z>$ (kpc)       &0.75&1.37  &2.41 &3.57 &4.50&5.53&6.42&7.42&8.30&9.75\\
\hline 
             & $<[Fe/H]>$ (dex)  &-0.31&-0.41&-0.76&-0.84&-1.19&-1.32&-1.51&-1.53&1.50&-1.30\\
\hline
             & m.e.              &$\pm$0.16&$\pm$0.25&$\pm$0.18&$\pm$0.12&$\pm$0.10& $\pm$0.10&
                                 $\pm$0.19&$\pm$0.16&$\pm$0.09&$\pm$0.08\\
\hline
\end{tabular}
\end{table*}

\section{Summary and discussion}

In this work we illustrated the capabilities of present and forthcoming analyses of 
CCD star-count data, when such analyses are based purely on star by star inversion 
of colour data, through  stellar photometric parallax. We showed how such analyses 
can be robust, provided that they utilise as a constraint consistency with the 
local solar neighbourhood stellar luminosity function. We showed how such analyses 
can limit possible metallicity gradients for the components of the Galaxy, and 
provide the choice of best model parameters. We now review our results and discuss 
them in the context of those by other authors.\\

{\it (i) The use of colour data to identify and reject extragalactic objects.}\\

A considerable fraction of the star candidates of Hall et al.\ (1996), selected from 
image structure, and labelled with "s", "sf", and "Fs" in their work turned out to 
be extra-galactic objects, according to their position in the $(U-B)_{o}$ - 
$(B-V)_{o}$ colour-colour diagram. An effective colour cut, consistent with those
adopted by {\it {SDSS}} (Chen et al.\ 2001) is to reject all point-sources with 
$(U-B)_{o}$ colour indices less than -0.46, which  corresponds to $(u'-g')_{o}<-0.50$.
Comparison of Fig.1 and Fig.2 shows that this single selection substantially reduces 
scatter away from the stellar locus. Removal of objects with $(U-B)_{o}<-0.46$ and 
imposing an apparent magnitude cut at the completeness limit $V_{o}\leq 20.5$ 
allowed the stellar locus to be readily identified, and outliers to be excluded.\\

{\it (ii) Stellar luminosity function at $r=0$ kpc obtained from deep CCD-photometry.}\\

The sample of Hall et al.\ (1996) does not allow space density determination for 
nearby stars due to the lack of apparently bright stars in this sample. Hence, space 
densities are complete at distances larger than 2.51, 1.00, and 0.63 kpc for absolute 
magnitude intervals $4<M(V)\leq5$, $5<M(V)\leq6$, and $6<M(V)\leq7$, respectively, 
and 0.40 kpc for four absolutely fainter intervals, i.e.: $7<M(V)\leq8$, $8<M(V)\leq9$, 
$9<M(V)\leq10$, and $10<M(V)\leq11$.

In order to allow comparison with the local luminosity functions of Hipparcos 
(Jahreiss \& Wielen\ 1997) and with that evaluated by Kul\ (1994) from the data of 
Gliese \& Jahreiss\ (1992), the star count models themselves must be used to 
extrapolate the star counts to the solar neighbourhood. While in general all three 
models analysed are in tolerable agreement with the required local normalizations,
there are differences. The model of Buser et al. fits best for three absolute magnitude 
intervals, i.e.: $6<M(V)\leq7$, $8<M(V)\leq9$, and $9<M(V)\leq10$, relative to that 
of Gilmore \& Wyse\ (Fig.9a). This difference is due to the differences between 
scale-heights and local densities adopted for the thick disk (cf Table 3). The model 
of Chen et al. matches the constraint well for low luminosity local thin disk stars, 
absolutely faint magnitude intervals ($M(V)>8$) whereas the model of Buser et al. 
matches better for the more luminous thick disk stars, for the bright ones ($M(V)<7$). 
This distinction is due to differences between five model parameters (Table 3). 
Additionally, one must take into account the difference density laws used for the halo
for these models, i.e.: de Vauceuleurs spheroid for Buser et al., and power-law for 
Chen et al.\

We determined the sensitivity of the local luminosity function constraint on 
determination of the axis ratio of the halo by calculating models following Chen et al. 
except with axis ratio $\eta=0.65$ (Fig.10a) and $\eta=0.84$ (Fig.10b). The first value 
(0.65) is that derived by Yanny et al.\ (2000) based on BHB tracers from {\it{SDSS}}
data, and rather close to the value (0.6) suggested by Wyse \& Gilmore\ (1988). The 
second value is not only equal or close to the values propesed by Buser et al. and 
Gilmore \& Wyse, but also it coincides with those cited by other authors. For example 
Hawkins\ (1984), and Bahcall \& Soneira\ (1984) found $\eta= 0.9$, and $\eta=0.8$
respectively. Preston et al.\ (1991) state that $\eta$ increases from $0.5$ to $1$ up 
to 20 kpc, while Robin et al.\ (2000) deduced that the halo has a flattening of 
$\eta=0.76$. It is interesting that the luminosity function comparison in figures 
10a and 10b distinguish these models, thus showing that the flattening parameter 
$\eta$ of the halo is the most sensitive parameter which can be distinguished here 
between the Buser et al. and the Chen et al. models. Finally, a de Vaucouleurs 
spheroid with the model parameters of Chen et al., except that $\eta=0.84$, works
well, (Fig.10c) indicating that $\eta$ but not the density law for the halo
plays an important role in the luminosity function comparison. Overall, we conclude 
that the model of Chen et al. (2001) is consistent with these data, under the 
condition that $\eta=0.84$.\\
\\
\\
\\
\\
{\it (iii) Vertical metallicity gradient for the three components of the Galaxy}\\

Our data are consistent with, but do not require, weak vertical metallicity gradients 
in both  the thin disk and thick disk. In the halo, any vertical metallicity gradient 
is even weaker. A better description of our data is that the metallicity distribution 
function is the sum of three discrete distributions, none of which has a significant 
metallicity gradient. Rather, an apparent vertical metallicity gradient arises from the 
changing contributions of the three distributions with distance from the Galactic Plane. 
Some gradient inside each population is however allowed by our analysis.

The maximum possible vertical metallicity gradient for the thin disk, i.e.:
$d[Fe/H]/dz \sim -0.2$ dex/kpc, is consistent with many other determinations, and 
consistent with a convolution of a weak age-metallicity relation and age-velocity 
dispersion relation.

If there were a detected vertical metallicity gradient for the thick disk this would 
impact some formation histories postulated for the formation of the classical thick disk. 
Until recently this component of our Galaxy was assumed to have a mean metal-abundance 
$[Fe/H]\sim-0.60$ dex, with a narrow metallicity range, a scale-height $1.0-1.3$ kpc, 
and that it comprises some $0.02-0.05$ of the material in the solar neighbourhood. 
Additionally, and more important, it was argued that the stars of thick disk were 
formed from a merger into the Galaxy (cf. Norris\ 1996 and references within), a 
formation mechanism unlikely to leave an abundance gradient. Some recent analyses
suggested that the thick disk is a more massive component of the Galaxy, 
(Majewski\ 1993) with a metal-poor (Norris\ 1996) and a metal-rich tail (Carney\ 2000; 
Karaali et al.\ 2000). Hence, a revision of the formation scenario of the thick disk 
may be required. The work of Reid \& Majewski\ (1993) in which a vertical metallicity 
gradient $d[Fe/H]/dz\sim-0.10$ dex/kpc is claimed is consistent with our results 
(Fig.12) but also consistent with a simple no-gradient mixed-population model. 
Chiba \& Yoshii\ (1998) also suggest a vertical metallicity gradient for the thick disk. 
A substantially larger sample of stars with both metallicities and appropriate kinematics 
will be required to distinguish between these models (cf Gilmore, Wyse, \& Norris 2002).

Detection of a metallicity gradient in the halo which changes with Galactocentric 
distance would be a test of scenarios suggesting important late accretion of the 
outermost part of the Galaxy: One might expect a gradient in the inner 
partly-dissipatively formed halo, and none farther out, provided that
the stellar velocity ellipsoid is as observed, only slightly radially
anisotropic. This gradient is consistent with 
our results: i.e.: there is a slight vertical metallicity gradient, 
$d[Fe/H]/dz\sim-0.10$ dex/kpc, in the inner part of the halo ($5<z\leq8$ kpc) and zero in
its outer part ($8<z\leq10$ kpc). However, we recognize that there are significant 
statistical uncertaintes and a proper interpretation will need to await large-scale 
stellar surveys from the {\it SDSS\/}.

\section*{Acknowledgments}
This work was supported by the Research Fund of the University of Istanbul. 
Project number 1417/050500. We thank to Hall et al.\ (1996) for providing 
their data.

\appendix

\section[]{The new metallicity calibration}

Data for 88 dwarfs with metallicities $-2.7\leq[Fe/H]\leq+0.26$ dex were
taken from three sources for a new metallicity calibration: (1) 57 of them
with $log g\geq4.5$ are from Cayrel de Strobel et al.\ (2001), a catalogue
which supplies detailed information for stars with abundance determinations
based on high-resolution spectroscopy. (2) 11 high or intermediate mass stars 
were taken from a different catalogue of the same authors (Cayrel de Strobel et
al.\ 1997). This catalogue has the advantage of including metal-poor stars
down to $[Fe/H]=-2.70$ dex with smaller gravity, i.e.: $log g\geq4.0$,
however. For the {\it UBV\/}-magnitudes and colours, specialised catalogues 
which are included in the General Catalogue of Photometric Data (Mermilliod et
al.\ 1997) were consulted. The parallax and the galactic latitude of stars
which were used in the choice of  the sample stars were provided from the 
database. (3) 20 stars classified as dwarfs by Carney\ (1979) who used them in 
his metallicity calibration were included also in the new sample.

The full interval for normalized ultra-violet excess, $-0.09\leq\delta_{0.6}\leq
+0.38$ mag was divided into 17 sub-intervals. The centroid of each was adopted 
as a locus point to fit the couple ($\delta_{0.6}, [Fe/H]$). Table A1
gives the locus points and the number of stars associated, and Fig.A1
the fit of these points by a third-degree polynomial, i.e.: $[Fe/H]=0.10-
2.76\delta-24.04\delta^{2}+30.00\delta^{3}$. Analysis of the deviations of
metallicities deduced from this calibration compared to the original metallicity 
shows that the accuracy is at the level of Carney\ (1979)'s work (Fig.A2a-c).

\begin{table}
\centering
\caption{Locus points and the number of stars associated with them (last
 column). The other colums give the current number, $\delta_{0.6}, [Fe/H]$, 
 mean errors for the $\delta_{0.6}$ and $[Fe/H]$ and, respectively.}
\begin{tabular}{cccccc}
\hline
No & $\delta_{0.6}$ & $[Fe/H]$ & $\Delta\delta_{0.6}$ & $\Delta[Fe/H]$ & N\\
\hline
~1 & $-$0.07 & +0.21 & 0.01 & 0.04 & 3 \\
~2 & $-$0.02 & +0.09 & 0.00 & 0.04 & 8 \\
~3 & +0.01 & +0.05 & 0.00 & 0.02 & 7 \\
~4 & +0.02 & +0.01 & 0.00 & 0.04 & 7 \\
~5 & +0.04 & $-$0.04 & 0.00 & 0.03 & 7 \\
~6 & +0.08 & $-$0.28 & 0.00 & 0.03 & 8 \\
~7 & +0.11 & $-$0.41 & 0.00 & 0.03 & 7 \\
~8 & +0.14 & $-$0.62 & 0.00 & 0.04 & 8 \\
~9 & +0.15 & $-$0.75 & 0.00 & 0.03 & 5 \\
10 & +0.17 & $-$0.93 & 0.00 & 0.04 & 4 \\
11 & +0.19 & $-$1.05 & 0.00 & 0.07 & 3 \\
12 & +0.22 & $-$1.32 & 0.00 & 0.04 & 5 \\
13 & +0.23 & $-$1.52 & 0.00 & 0.06 & 3 \\
14 & +0.26 & $-$1.68 & 0.00 & 0.03 & 3 \\
15 & +0.28 & $-$2.05 & 0.00 & 0.06 & 4 \\
16 & +0.31 & $-$2.10 & 0.00 & 0.04 & 3 \\
17 & +0.36 & $-$2.60 & 0.01 & 0.05 & 3 \\
\hline
\end{tabular}
\end{table}

\begin{figure}
\caption{The third-degree polynomial curve throught 17 locus-points and the
correlation coefficient. The bars show the mean errors.}
\end{figure}

\begin{figure}
\caption{Deviation of evaluated metallicities from original ones versus
original metallicity for all stars in our sample (a), for stars with
$[Fe/H]\geq-1.75$ dex in our sample (b), and for the sample of Carney (c),
where $[Fe/H]=-1.75$ dex is the validity limit for the Carney\ (1979)
calibration.}
\end{figure}

\bsp

\label{lastpage}

\end{document}